\documentstyle[prd,aps,preprint,tighten]{revtex}

\begin{document}
\draft
\title{Magnetic moments of the 3/2 resonances and their quark spin
structure}

\author{Johan Linde\footnote{Electronic address: jl@theophys.kth.se}
and H{\aa}kan Snellman\footnote{Electronic address: snell@theophys.kth.se}}
\address{Department of Theoretical Physics\\
Royal Institute of Technology\\
S-100 44 STOCKHOLM\\
SWEDEN}

\date{\today}

\maketitle

\begin{abstract}
We discuss magnetic moments of the $J=3/2$ baryons based on an earlier
model for the baryon magnetic moments, allowing for flavor symmetry breaking
in the quark magnetic moments as well as a general quark spin
structure. From our earlier analysis of the nucleon-hyperon
magnetic moments and
the measured values of the magnetic moments of  $\Delta^{++}$ and
 $\Omega^{-}$ we predict the other
magnetic moments and deduce the spin structure of the resonance particles.
We find from experiment that the total spin polarization of the decuplet
baryons, $\Delta\Sigma(3/2)$, is considerably
smaller than the non-relativistic quark model
value of 3, although the data is still not good
enough to give a precise determination.
\end{abstract}


\narrowtext

\section{Introduction}
The recent precise measurement of the magnetic moment of $\Omega^{-}$,
$\mu(\Omega^{-})=-2.024\pm 0.056$~$\mu_N$\cite{omega},
gives new possibilities for theoretical
studies of the magnetic moments of the decuplet baryons.
These magnetic moments constitute an important domain for investigating baryon
structure and have earlier been studied {\it e.g.\/} in quenched lattice gauge
theory\cite{draper}, quark models\cite{schlumpf}, the chiral bag
model\cite{Brown} and chiral perturbation theory\cite{butler}.

In an earlier paper\cite{jlhs} we introduced
a general parameterization of the baryon
magnetic moments in order to account for necessary modifications of the
non-relativistic quark model (NQM) description. In this paper we
extend this parameterization to the $J=3/2$ resonance particles in the
decuplet.

Although only two of the magnetic moments of these states have been
measured,  it is still in principle
possible to obtain interesting information on
the spin structure of the resonance particles from these data. We analyze
the magnetic moments of the spin $3/2$ particles in this model and
make predictions for the magnetic moments of the resonance particles using
values for the quark magnetic moments extrapolated from the nucleon-hyperon
data in our previous analysis.

\section{Analysis of magnetic moments for the spin 3/2 resonances}
The magnetic moment of a hadron in isomultiplet $B$ can,
in the quark model, be written as a linear sum of
contributions from the various flavors
\begin{equation}
        \mu(B^{i}) = \mu_{u}\Delta u^{B^{i}} +
        \mu_{d}\Delta d^{B^{i}} + \mu_{s}\Delta
        s^{B^{i}},
\end{equation}
where $\mu_{f}$ is an effective magnetic moment
of the quark of flavor $f$ and $\Delta f^{B^{i}}$
is the corresponding spin polarization for baryon $B^{i}$, $i$ being the
baryon charge state. By symmetry arguments the $\Delta f^{B^{i}}$'s in the
octet baryons can be expressed as constant linear
combinations of the three $\Delta f$'s for the proton, which are the only
spin polarizations needed to describe the octet:
\begin{equation}
        \Delta f^{B^{i}} = \sum_{f'}^{}M(B^{i})_{ff'}\Delta f',
\end{equation}
where $f,f'$ runs over $u,d,s$, and the $M(B^{i})$'s are matrices with
constant elements. In particular, for the six mirror symmetric
baryons of type $B(xyy)$, where $x$ and $y$ are different flavors, we have
$\Delta y^{B^{i}}=\Delta u$, $\Delta x^{B^{i}}=\Delta d$ and $\Delta
z^{B^{i}}=\Delta s$, where the flavor $z$ is the non-valence quark flavor.
In the NQM the values of these spin
polarizations are $\Delta u = \frac{4}{3}$, $\Delta
d = -\frac{1}{3}$ and $\Delta s = 0.$

In the model we consider in Ref.\cite{jlhs},
we let the quark magnetic moments be
different in different baryon isomultiplets, but the
same within each isospin multiplet\cite{avenarius}.
When the quark magnetic moments are taken to be the same in all isomultiplets,
the baryon magnetic moments are
connected by sum-rules. These sum-rules are
clearly violated by the experimental data indicating that the quark
magnetic moments are not the same in different isomultiplets.
Lattice calculations\cite{leinweber} also indicate that
they are different in different baryons. Thus, the quark magnetic moments
will be denoted by $\mu_f^B$, where $f$ is the quark flavor and $B$ is
the baryon isomultiplet.

The spin
structure variables in the magnetic moments are
not a priori the same as in deep inelastic
scattering experiments and  axial-vector form factors. However, in many
models\cite{karl} they are proportional to these spin polarizations.
Since the equations
for the magnetic moments are homogeneous in the quark magnetic moments and
the spin polarizations, the quark magnetic moments can be deduced from
experimental data by normalizing the spin polarizations with the value of
the weak axial-vector form factor
$g_{A}^{np}=\Delta u-\Delta d=1.2573$\cite{data}. The spin
polarizations we deduce will then be the relevant
ones for the spin as measured in deep
inelastic scattering experiments. Later we will
use data of the quark magnetic moments from the octet
baryons, and these are normalized in this way.

In the decuplet there are four different isomultiplets and thus
there are in principle twelve quark magnetic moments. We
introduce the flavor breaking parameters $T$ and $U$ describing the
flavor breaking among the quarks, which is assumed to be the same for all
isomultiplets. Thus $\mu_u^B = T \mu_d^B$ and $\mu_s^B = U \mu_d^B$
independently of $B$. In what follows we will assume that $T$ and $U$ are
the same for both the octet and decuplet
baryons.\footnote{In the NQM the values of
$T$ and $U$ are $T=\mu_u/\mu_d = -1.91,
U=\mu_s/\mu_d = 0.63$ when the three quark magnetic moments are
allowed to be free parameters.} This assumption can in principle be tested
when more data of the decuplet magnetic moments is available.
There are therefore six parameters for the twelve
quark magnetic moments.

In each
baryon there are also spin structure variables. In the octet there are
three different ones, $\Delta u,\Delta d $
and $\Delta s$, defined as the spin polarization of the $u$, $d$ and $s$
quarks in the proton in the spin up state.
However, in the decuplet it turns out that there
are only two different spin polarization parameters,
 denoted $\Delta u$ and $\Delta s$. These are both
defined for the $\Delta^{++}$ resonance in the $J_z=+3/2$ state and
are normalized for convenience of notation with a factor of three,
so that $3\Delta u$, $3\Delta s$ and $3\Delta s$
are the quark spin
polarizations
in $\Delta^{++}$  of the $u$, $d$ and $s$ quarks respectively.
The reason for there being only two different
spin polarization parameters in the decuplet is
that the spin structure there is much simpler than
in the octet due to the flavor functions being
fully symmetric, which does not allow mixed symmetry for the spins.

 In the
NQM the spin polarization parameters have for the decuplet the values
$\Delta u =1 $ and
$\Delta s = 0$. In a more sophisticated model,
due to virtual quark antiquark pairs, gluonic corrections,
spontaneously broken chiral symmetry, emission and capture of Goldstone
bosons etc.\ they might have more general values.

SU(3) relations are used to construct the spin structure of the
states. The flavor breaking is  assumed to occur in the quark
magnetic moment operators by the effective
masses and possible vertex corrections.
The magnetic moments of
the decuplet particles can then be written:
\begin{mathletters}
\begin{eqnarray}
\mu(\Delta^{++})= & &\mu_d^{\Delta}   (3T\Delta u   +  (3+3U) \Delta s),\\
\mu(\Delta^{+})= &&\mu_d^{\Delta}    ((2T+1)\Delta u+  (T+2+3U)\Delta s),
\label{tank}\\
\mu(\Delta^{0})= &&\mu_d^{\Delta}    ((T+2)\Delta u +  (2T+1+3U)\Delta
s),
\label{sa}\\
\mu(\Delta^{-})= &&\mu_d^{\Delta}    (3\Delta u+       (3T+3U) \Delta s),\\
\mu(\Sigma^{*+})=&&\mu_d^{\Sigma^{*}}((2T+U)\Delta u +  (T+3+2U)\Delta
s) ,
\label{kul}\\
\mu(\Sigma^{*0})=&&\mu_d^{\Sigma^{*}}((T+1+U)\Delta u + (2T+2+2U)\Delta
s),\\
\mu(\Sigma^{*-})=&&\mu_d^{\Sigma^{*}}((2+U)\Delta u  +  (3T+1+2U)\Delta
s),
 \label{vi}\\
\mu(\Xi^{*0})   =&&\mu_d^{\Xi^{*}}       ((2U+T)\Delta u+ (2T+3+U)\Delta s),
 \label{har}\\
\mu(\Xi^{*-})   =&&\mu_d^{\Xi^{*}}       ((2U+1)\Delta u+ (3T+2+U)\Delta
s),
 \label{hurra}\\
\mu(\Omega^{-}) =&&\mu_d^{\Omega}    (3U\Delta u+    (3+3T) \Delta s).
\end{eqnarray}
\end{mathletters}

Combining the
equations (\ref{tank},\ref{sa},\ref{kul},\ref{vi},\ref{har} and \ref{hurra})
we get the system of equations
\widetext
\begin{mathletters}\label{egenekv}
\begin{eqnarray}
(3(T+1)-A'(T-1))\Delta u    + (3(T+1+2U)+A'(T-1))\Delta s=&&0, \label{egen1}\\
(T+1+U-B'(T-1))\Delta u + (2(T+1+U)+B'(T-1))\Delta s=&&0, \label{egen2}\\
(1+T+4U -C'(T-1))\Delta u   +  (5T+5+2U+C'(T-1)) \Delta s=&&0, \label{egen3}
\end{eqnarray}
\end{mathletters}
\narrowtext
where
\begin{eqnarray}
A'=&&\frac{\mu(\Delta^{+})+\mu(\Delta^{0})}{\mu(\Delta^{+})-\mu(\Delta^{0})},\\
B'=&&\frac{\mu(\Sigma^{*+})+\mu(\Sigma^{*-})}{\mu(\Sigma^{*+})-
\mu(\Sigma^{*-})},\\
C'=&&\frac{\mu(\Xi^{*0})+\mu(\Xi^{*-})}{\mu(\Xi^{*0})-\mu(\Xi^{*-})}.
\end{eqnarray}

The set of equations (\ref{egenekv}) constitutes an over-determined system.
 We therefore get two
different equations from the conditions that the secular determinant of either
(\ref{egen1})+(\ref{egen2}) or (\ref{egen2})+(\ref{egen3})  should vanish.
Both of these contain the
root $U=-1-T$, which we discard. The other roots are
\begin{eqnarray}
        U & = & \frac{1}{2}(1+T+(A'-2B')(1-T)),  \\
        U & = & \frac{1}{2}(1+T+(2B'-C')(1-T)).
\end{eqnarray}

 A similar analysis for the octet baryons gives the relation\cite{jlhs}
 \begin{equation}
        U=\frac{1}{2}(1+T+D(1-T)). \label{UT-relation}
 \end{equation}
 where $D=\sqrt{AB-AC+BC}=0.78\pm 0.02$~$\mu_N$, and
 \begin{eqnarray}
        A= && \frac{\mu(p)+\mu(n)}{\mu(p)-\mu(n)}, \label{A} \\
        B= &&
        \frac{\mu(\Sigma^{+})+\mu(\Sigma^{-})}{\mu(\Sigma^{+})-
        \mu(\Sigma^{-})} , \\
        C= && \frac{\mu(\Xi^{0})+\mu(\Xi^{-})}{\mu(\Xi^{0})-\mu(\Xi^{-})}  .
        \label{C}
\end{eqnarray}
  Our
 assumption that the symmetry breaking is the same in the decuplet as in
 the octet predicts the sum-rules
 \begin{equation}
        A'-2B'=2B'-C'=0.78\pm 0.02\mu_N.
 \end{equation}

In the present case there are two more baryons than in the
nucleon-hyperon case and one spin polarization less. It might therefore seem
possible to get the spin structure directly from the magnetic moment data.
However, as we mentioned above, the equations
are homogeneous in the quark moments and the spin polarizations. We
therefore still lack a normalization for the spin polarizations.
 Hence, the set of equations above
can only determine the relative magnitude of the spin polarizations
and quark magnetic moments. Also the magnetic moment of $\Omega^{-}$
cannot be given in the model, since the value of $\mu_{d}^{\Omega}$ is a free
parameter. We therefore have nine data with seven parameters, constituting
an over-constrained system, giving the possibility of a non-trivial test of
the model.

\widetext
For example using only the $\Delta$-resonances we can solve for the spin
polarization in terms of $\mu_{d}^{\Delta}$. The relevant formulas are
\begin{eqnarray}
        3\Delta u & =
        & \frac{1}{\mu_{d}^{\Delta}(T-1)(T+U+1)}[(2T+1+3U)\mu(\Delta^{+})-
        (T+2+3U)\mu(\Delta^{0})] ,\\
         3\Delta s & = &
         \frac{1}{\mu_{d}^{\Delta}(T-1)(T+U+1)}
         [-(T+2)\mu(\Delta^{+})+(2T+1)\mu(\Delta^{0})].
\end{eqnarray}

The total spin polarization, $\Delta\Sigma(3/2)$,
is given by the sum of the coefficients of the
quark magnetic moments, {\it i.e.\/} in our model of $\mu_{d}^{B}$,
$T\mu_{d}^{B}$ and $U\mu_{d}^{B}$.
This gives for all 3/2 baryons $\Delta \Sigma(3/2)=3(\Delta u+2\Delta s)$. Thus
\begin{equation}
        \Delta \Sigma(3/2) =
\frac{3}{\mu_d^{\Delta}(T-1)(T+U+1)} [(U-1)\mu(\Delta^{+}) +
(T-U)\mu(\Delta^{0})] .
\end{equation}
However, we can also express the spin sum in terms of two other of these
moments as
\begin{equation}
                \Delta \Sigma(3/2) =
                \frac{3}{\mu_d^{\Delta}(T-1)(T+U+1)}[(T+U-2)\mu(\Delta^{++})+
                (2T-U-1)\mu(\Delta^{-})].
\end{equation}
We will not pursue this line further here, since there is at present not
enough data to test these relations.

\narrowtext
\section{Predicting decuplet magnetic moments and
spin polarizations from the octet data}
Lacking most of the data to carry out an analysis,
we will from now on instead work
to predict the magnetic
moments of the $3/2$ resonances using data from the nucleon-hyperon
system. We will then also be able to obtain the quark spin
polarizations of the spin $3/2$ resonances.

As we mentioned above, the flavor symmetry breaking is assumed to be the
same for the decuplet as for the octet. This assumption
can in the future be tested in
the way indicated above, by determining
the constants $T$ and $U$ from the magnetic
moments in the decuplet.

{}From the measured magnetic moments of the octet
baryons we can calculate the $\mu_{d}^{B}$'s, once the spin polarizations
are known. Since the value of $\Delta\Sigma$
for the octet (the nucleons) is still not
too well determined, and does not give too precise
a determination of $T$, we will take two
typical cases, one consistent with the most recent data, and for the other
we take $T=-2$ corresponding
to no isospin symmetry breaking. New evaluations of $\Delta\Sigma$ for the
nucleons give different, although overlapping values. A recent
determination by SMC\cite{SMC} gives $\Delta\Sigma = 0.20\pm0.11$, whereas
Ellis and Karliner\cite{Ellis} have analyzed data to
get $\Delta\Sigma = 0.31\pm 0.07$.

The values we will use are
$\Delta\Sigma= 0.27 $ corresponding to $T= -1.80$ and
$\Delta\Sigma = 0.08 $ for $T=-2$ (no isospin symmetry breaking).
The formulas used to extract the $\mu_{d}^{B}$'s are\cite{jlhs}
\begin{eqnarray}
        \mu_{d}^{N}= &  & \frac{\mu(p)-\mu(n)}{(T-1)g^{np}_{A}},    \\
    \mu_{d}^{\Lambda}= &  & \frac{6
    \mu(\Lambda)}{(T-1)g^{np}_{A}}\frac{B-C}{2AB-2AC+2CD-CB-BD},\\
        \mu_{d}^{\Sigma}= &  &
        \frac{\mu(\Sigma^{+})-\mu(\Sigma^{-})}{(T-1)g^{np}_{A}}
        \frac{B-C}{D-C},  \\
         \mu_{d}^{\Xi}=&  &
         \frac{\mu(\Xi^{0})-\mu(\Xi^{-})}{(T-1)g^{np}_{A}}\frac{B-C}{D-B},
\end{eqnarray}
where  $g^{np}_{A}=1.2573$ and  $A$, $B$ and $C$ are given by
(\ref{A}--\ref{C}).

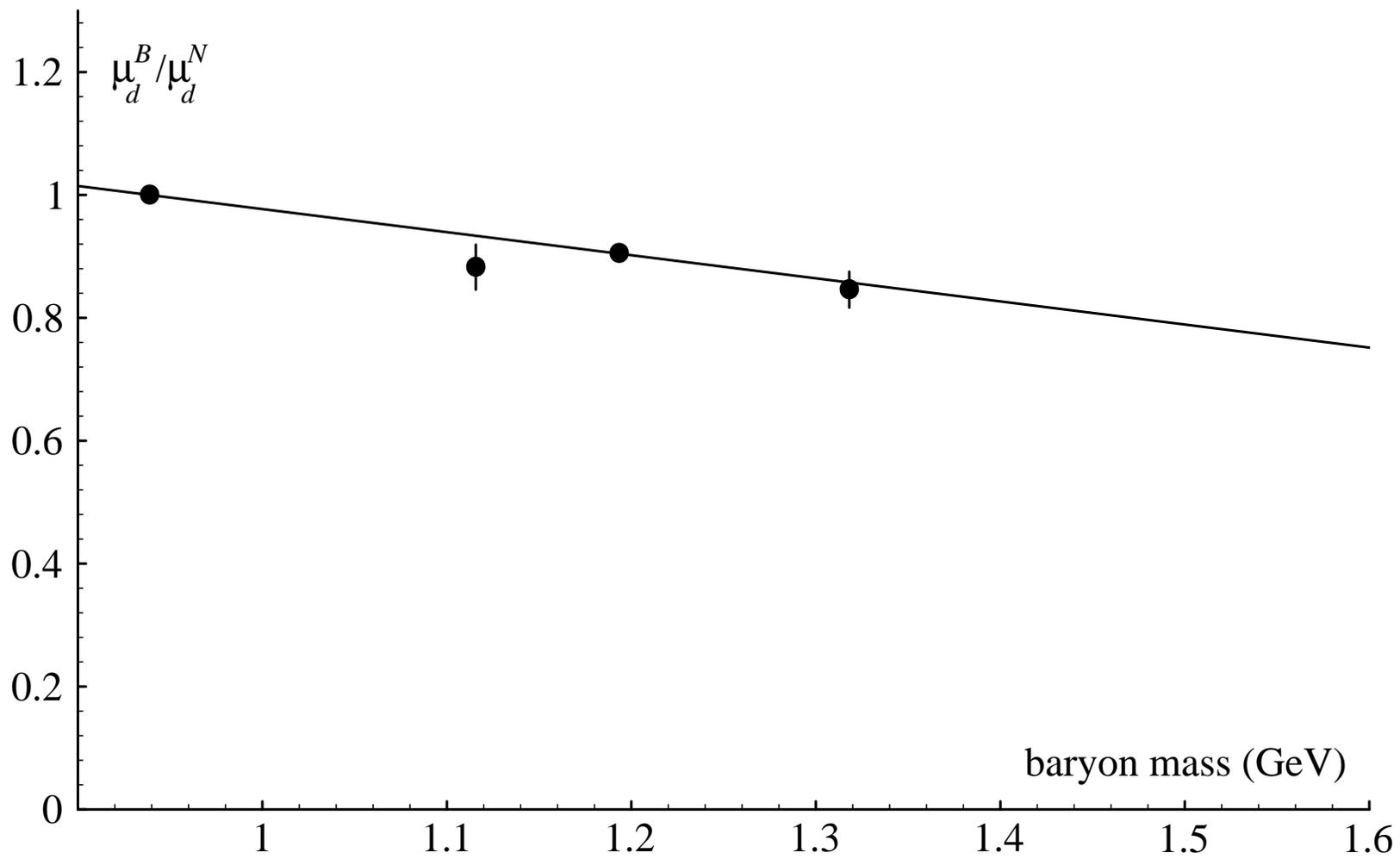
\begin{figure}
\caption{$\mu_d^{B}/\mu_d^{N}$ as a function of the  mass of  $B$.
The straight line is a least square fit to the four data points.
If the line is extrapolated
to $\Omega^{-}$ all $\mu_d^{B}$'s for the decuplet can be found.}
\label{vacker_linje}
\end{figure}

In Fig.~\ref{vacker_linje} the ratios $\mu_d^{B}/\mu_d^{N}$
are plotted as a function of
the mean mass of $B$. We see that they can be well fitted by a linear
function, and we therefore extrapolate
this function out to $\Omega^{-}$ and use
the ensuing data set for the
$\mu_{d}^{B}$'s, where the intermediate values are interpolated from this
graph.
These values are given in Table \ref{tabell1}.
We will analyze this linear relation among the $\mu_d^{B}$'s in
Section~\ref{snackmotslutet}.

With the parameters of Table \ref{tabell1} we can now determine
the quark spin polarizations in the
decuplet from the measured magnetic moments of $\Delta^{++}$ and
$\Omega^{-}$ and use this to predict the other magnetic moments.

The result is

\begin{eqnarray}
        3\Delta u  & = & \frac{\mu_{d}^{\Omega}(1+T)\mu(\Delta^{++})
        -\mu_{d}^{\Delta}(1+U)\mu(\Omega^{-})}{\mu_{d}^{\Delta}
        \mu_{d}^{\Omega}(T(T+1)-U(U+1))},
        \label{delta-u} \\
3\Delta s & = &  \frac{-\mu_{d}^{\Omega}U\mu(\Delta^{++})+
\mu_{d}^{\Delta}T\mu(\Omega^{-}) }
{\mu_{d}^{\Delta}\mu_{d}^{\Omega}(T(T+1)-U(U+1))}.
        \label{delta-s}
\end{eqnarray}

Inserting the experimental values given in Table \ref{tabell2} gives
\begin{eqnarray}
        3\Delta u  & = & 0.4 \pm 1.1, \\
        3\Delta s & = & -2.0 \pm 0.8, \\
        \Delta\Sigma(3/2)=3(\Delta u +2\Delta s) & = & -3.6 \pm 2.6,
\end{eqnarray}
for $T=-2$ and
\begin{eqnarray}
        3\Delta u & = & -1.9 \pm 2.8, \\
       3\Delta s & = & -4.3\pm 2.5, \\
        \Delta\Sigma(3/2)=3(\Delta u+2\Delta s) & = & -10.5 \pm 7.9,
\end{eqnarray}
for $T=-1.80$.

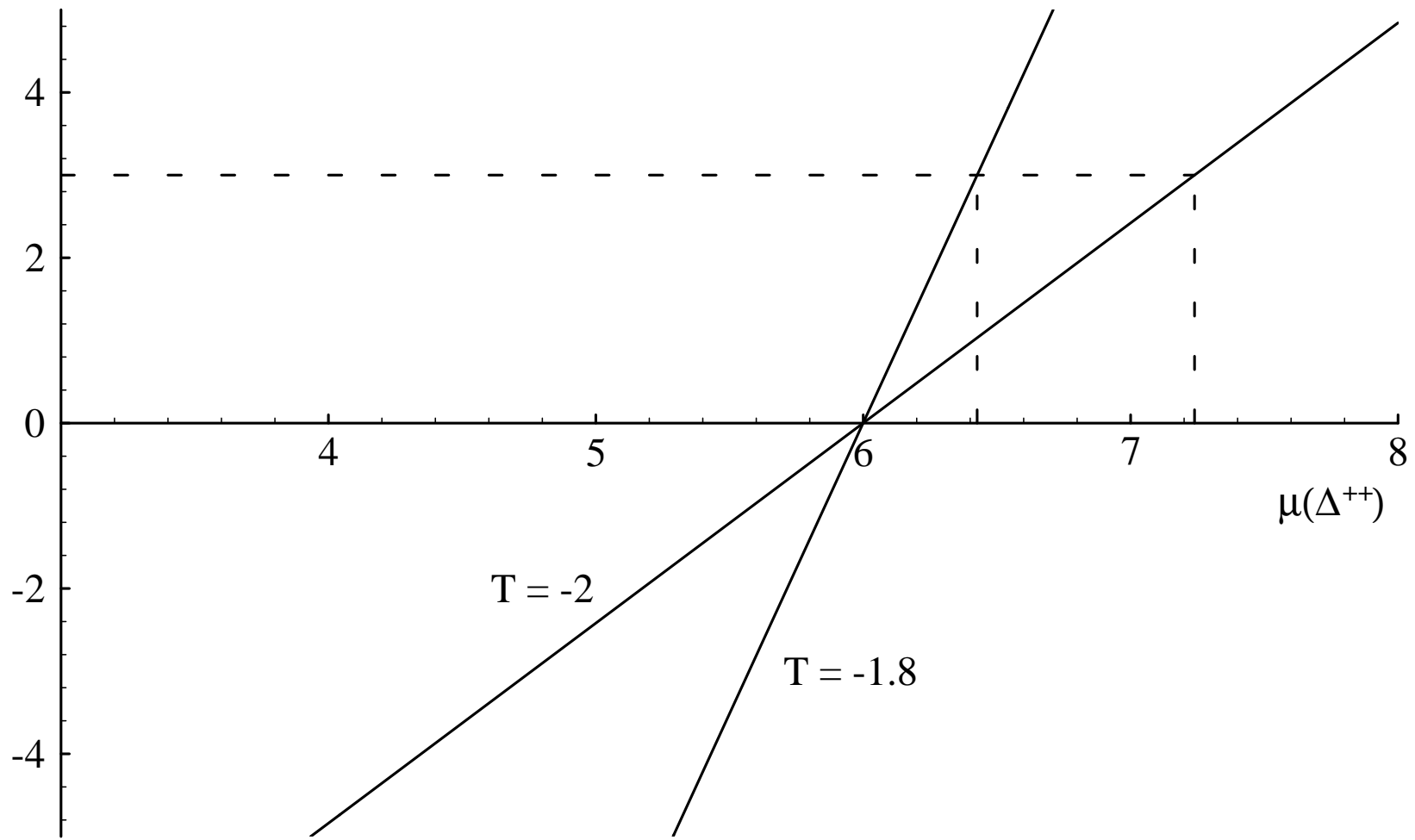
\begin{figure}
\caption{$\Delta\Sigma(3/2)$ as a function of $\mu(\Delta^{++})$ for
$T=-2$ and $T=-1.8$. We see that $\Delta\Sigma(3/2)=0$ for
$\mu(\Delta^{++})=6.0$~$\mu_N$ and that $\Delta\Sigma(3/2)=3$ for
$\mu(\Delta^{++})=7.2$~$\mu_N$ when $T=-2$,
and for $\mu(\Delta^{++})=6.4$~$\mu_N$
when $T=-1.8$. $\Delta\Sigma(3/2)$ depends very strongly on $T$, and
therefore no precise prediction can be made.} \label{tjusig_kurva}
\end{figure}

The result indicates that the decuplet spin polarization with the present
data taken at face value is large and negative.
However, the error on the magnetic moment of
$\Delta^{++}$ is rather large and the value
might according to Ref.\cite{data} be anywhere
between $3.7$ and $7.5$~$\mu_N$. We have therefore instead plotted
the value of $\Delta\Sigma(3/2)$ as
a function of $\mu(\Delta^{++})$ in
Fig.~\ref{tjusig_kurva}. From this we see that
$\Delta\Sigma(3/2)$ passes zero at around $6.0$~$\mu_N$
independently of the value
of $T$.  Only at $\mu(\Delta^{++})\approx
7.2$~$\mu_N$ does $\Delta\Sigma (3/2)$
reach the NQM value of $3$ when $T=-2$. For $T=-1.8$  $\Delta\Sigma (3/2)$
reaches the value 3 at $\mu(\Delta^{++})\approx 6.4$~$\mu_N$.

We clearly need a better measurement of $\mu(\Delta^{++})$ to truly fix
the value of $\Delta\Sigma(3/2)$, but
in face of existing data we feel confident to say that $\Delta\Sigma (3/2)$ is
much smaller than the NQM value of $3$, in analogy to the situation
for the nucleon spin.

For completeness we have included a prediction of the
magnetic moments of the decuplet assuming that $\mu(\Delta^{++})\approx
6.0$~$\mu_N$, corresponding to $\Delta\Sigma(3/2)=0$,
in Table \ref{tabell2}. This value also gives $3\Delta u=1.9$ and $3\Delta
s=-0.9$, when $T=-2$, which looks more realistic.

Using (\ref{delta-u}) and (\ref{delta-s}) the magnetic moments of the decuplet
particles can be predicted.
The values are given in Table \ref{tabell2}.
 We first remark that these values are
independent of $T$. To see this let us introduce the combinations
\begin{equation}
        \frac{\mu(\Delta^{++})}{3\mu_{d}^{\Delta}}= T \Delta u + (1+U) \Delta s
\end{equation}
and
\begin{equation}
        \frac{\mu(\Omega^{-})}{3\mu_{d}^{\Omega}} = U \Delta u +(1+T) \Delta s.
\end{equation}
In terms of these two numbers we can construct all other magnetic moments
as linear combinations of them. We have
\widetext
\begin{mathletters}
\begin{eqnarray}
        \mu(\Delta^{+})= && \frac{3D+1}{3(D+1)}\mu(\Delta^{++})+
        \frac{2}{3(D+1)} \frac{\mu_d^{\Delta}}{\mu_{d}^{\Omega}}
        \mu(\Omega^{-}), \\
        \mu(\Delta^{0})= && \frac{3D-1}{3(D+1)}\mu(\Delta^{++})
        +\frac{4}{3(D+1)} \frac{\mu_d^{\Delta}}{\mu_{d}^{\Omega}}
        \mu(\Omega^{-}), \\
        \mu(\Delta^{-}) = &&  \frac{D-1}{D+1}
                \mu(\Delta^{++})
                +\frac{2}{D+1} \frac{\mu_d^{\Delta}}{\mu_{d}^{\Omega}}
                \mu(\Omega^{-}), \\
        \mu(\Sigma^{*+})= &&
        \frac{2\mu_d^{\Sigma^{*}}}{3\mu_d^{\Delta}}\mu(\Delta^{++})
         + \frac{\mu_d^{\Sigma^{*}}}{3\mu_{d}^{\Omega}}\mu(\Omega^{-}), \\
        \mu(\Sigma^{*0})= && \frac{2D}{3(D+1)}
        \frac{\mu_d^{\Sigma^{*}}}{\mu_d^{\Delta}}\mu(\Delta^{++})
         + \frac{D+3}{3(D+1)}\frac{\mu_d^{\Sigma^{*}}}{\mu_{d}^{\Omega}}
         \mu(\Omega^{-}),  \\
        \mu(\Sigma^{*-})= && \frac{2D-2}{3(D+1)}
        \frac{\mu_d^{\Sigma^{*}}}{\mu_d^{\Delta}}\mu(\Delta^{++})
         + \frac{D+5}{3(D+1)}\frac{\mu_d^{\Sigma^{*}}}{\mu_{d}^{\Omega}}
         \mu(\Omega^{-}) , \\
        \mu(\Xi^{*0})= &&
        \frac{\mu_d^{\Xi^{*}}}{3\mu_d^{\Delta}}\mu(\Delta^{++})
         + \frac{2\mu_d^{\Xi^{*}}}{3\mu_{d}^{\Omega}}\mu(\Omega^{-}),   \\
        \mu(\Xi^{*-})= &&\frac{D-1}{3(D+1)}
        \frac{\mu_d^{\Xi^{*}}}{\mu_d^{\Delta}}\mu(\Delta^{++})
         +
         \frac{2D+4}{3(D+1)}\frac{\mu_d^{\Xi^{*}}}{\mu_{d}^{\Omega}}
         \mu(\Omega^{-}) ,
\end{eqnarray}
\end{mathletters}
\narrowtext
where $D =\sqrt{AB-AC+BC}$. We have here used the relation
(\ref{UT-relation}) between $U$ and $T$.  Since the ratios between the
$\mu_d^{B}$'s are independent of $T$, the baryon magnetic moments also
are.

The ratios between the $\mu_d^{B}$'s within the decuplet found from the
linear fit resembles very much those found in lattice gauge theory
calculations\cite{draper}.

The magnetic moments differ notably from
those of the NQM, mainly because of the
low experimental value of $\mu(\Delta^{++})$. But we also note that
the recent precise measurement of $\mu(\Omega^{-}) = -2.024\pm
0.056$~$\mu_N$\cite{omega} is off from the NQM value $3\mu_{s}=
-1.84$~$\mu_N$
by several standard deviations.

\section{The mass dependence of the $\mu_{d}$'s}\label{snackmotslutet}
Here we will study the linear relation among the $\mu_d^{B}$'s that we
found when analyzing the octet data in the model.

We first remark that the
quark magnetic moments obtained with this model are really only effective
moments, as was already mentioned in the introduction.

In the quark model the magnetic moment of a quark is of the form
\begin{equation}
        \mu_{f}=\frac{e_{f}}{2m_{f}},
        \label{next}
\end{equation}
$e_{f}$ being the quark charge. The dependence on $B$ could then be
due to a variation of the effective quark mass $m_{f}$ on $B$.
A model of this kind has been
considered by Chao\cite{chan}.

Another interpretation is suggested by a further study of equation (1).
Let us introduce the ratio
\begin{equation}
        \alpha(B) = \frac{\mu_{d}^{B}}{\mu_{d}^{N}},
\end{equation}
which as a function of the mean mass of $B$ is plotted in
Fig.~\ref{vacker_linje}. With this
function we can now rewrite equation (1) as follows:
\begin{equation}
                \mu(B^{i}) = \sum_{f,f'}\mu_{f}\alpha(B)
                M(B^{i})_{ff'}\Delta f' ,
                \label{BBB}
\end{equation}
where $\mu_{f}=\mu_{f}^{N}$, $f,f'=u,d,s$. Since the $\alpha(B)$'s are
independent of $f$ we can take them outside the sum to get
\begin{equation}
                \mu(B^{i}) = \alpha(B) \sum_{f,f'}
                \mu_{f}M(B^{i})_{ff'}\Delta f' .
\end{equation}

This equation suggests that the magnetic moments of heavier states for some
reason decrease with increasing mass relative to the values expected from
the nucleon data. This might take place through collective effects that
increase with mass, {\it e.g.\/} the
moment of inertia, and/or gluonic collective
effects etc. One example of this could be Skyrmion collective effects,
like in the chiral bag model\cite{Brown}.

Finally a third way of interpreting the result is also possible.

In our analysis the $\Delta f$'s are kept constant
and the same in the whole octet.
We then get $B$-dependent
$\mu_{d}$'s in the form
\begin{equation}
        \mu_{d}^{B} =\alpha(B)\mu_{d}^{N}.
\end{equation}
However, we also note that by a slight reinterpretation of our model we
could associate the $\alpha(B)$-factor instead with the $\Delta f$'s. In this
case the quark magnetic moments would be fixed throughout and the spin
polarizations would be varying linearly with the mass of $B$ in the form
\begin{equation}
        \Delta f^{B^{i}} = \sum_{f'}M(B^{i})_{ff'}\alpha(B)\Delta f'
\end{equation}
for each flavor $f$.

Since the baryon masses are well accounted for by mass formulas
with the same quark masses in both the octet and the decuplet, equation
(\ref{next}) suggests that the quark magnetic moments might also be constant in
the two supermultiplets. This last interpretation is therefore not
unreasonable. Our empirically
found mass dependence of the $\mu_{d}^{B}$'s would then be induced
by us keeping the $\Delta f$'s fixed in the analysis.

Clearly, as long as it comes to analyzing only the baryon magnetic
moments, these different interpretations of equation
(\ref{BBB}) are equivalent, and all relations
between magnetic moments will be the same in the three interpretations.

In the last interpretation, however, the quark spin
polarizations would decrease with increasing
isomultiplet mass. Thus $\Delta\Sigma = 0.27$ for
nucleons will decrease to $\approx 0.23$ for
the $\Xi$-particles. Also the spin  polarizations as
well as $\Delta\Sigma(3/2)$ should be renormalized from
the value presented here by a factor $\alpha(B)$
for the $B^{i}(3/2)$ resonances,
since we have normalized the spin polarization relative to the nucleons.
This means in particular that $\Delta\Sigma(3/2)$ for the
$\Delta$-resonances should instead have the value $\Delta\Sigma(3/2)= -3.2\pm
2.6$ for $T=-2$.
These questions could be illuminated if we could measure the spin
polarizations in the spin $3/2$ resonances.

In all interpretations we have assumed that
$\alpha(B)$ is valid for both the octet and decuplet. This implies that
the magnetic moments of the
baryons are predicted to be vanishingly small as $\alpha(B)$ goes to zero
as a function of the mass $m_{B}$ of $B$.
This happens at $m_{B}\approx 3.6$ GeV
if the linear relation is still valid in this region.
At any rate it is clearly highly
interesting to try to measure magnetic moments of very massive baryon
states and to study them with lattice calculations.

Further studies of the last interpretation will be presented
elsewhere\cite{jlhs3}.

In any case we can test the linear relation among the quark magnetic
moment ratios by the value of  $\mu_{d}^{\Omega}$
obtained from a fit to say $\mu(\Omega^{-})$ once the values of $\Delta u$
and $\Delta s$ are given from any other two magnetic moments measured in
the future.

\section{Summary and Conclusions}
We have studied the magnetic moments of the spin $3/2$ resonances in a
model allowing a general parameterization of flavor symmetry breaking and
arbitrary quark spin polarizations. As we saw
in an earlier analysis of the octet magnetic moments,  the flavor breaking
is small when the quark spin of the nucleon is small. At face value of the
measured magnetic moments, the isospin symmetric case $T=-2$ suggests a
negative quark spin content in the  $3/2$ resonances of the order of
$-3.6\pm 2.6$ as compared to the NQM value of $3$. However, the
experimental data is still not good enough to give a precise determination.

Using the two measured magnetic moments we predict the magnetic moments of
the other $3/2$ resonances independently of the symmetry breaking.
They are quite different from the
quark model ones (as are already the measured
values of $\mu(\Delta^{++})$ and $\mu(\Omega^{-})$),
and could, when measured,
be used as a probe to study the consistency of the model.
 In fact, already three experimental data is enough
to test most of the assumptions made in our model.

We also find that the quark magnetic
moment ratios in our analysis decrease linearly with mass of the host
particle and discuss various interpretations of this effect. It would be
very interesting to try to measure magnetic moments of very massive baryon
states and to study them with lattice calculations, to see how far this
linear dependence holds.

\acknowledgments

This work was supported
by the Swedish Natural Science Research
Council (NFR), contract F-AA/FU03281-308.

\newpage

\begin{table}
\caption{The parameters used in the model.
The decuplet quark magnetic moments are
interpolated from a linear fit to the octet ones.
The error of $0.05$~$\mu_N$ for
all decuplet moments is an estimate of the theoretical uncertainty. All
moments are given in $\mu_N$.} \label{tabell1}
\begin{tabular}{ldd}
&Case $T=-2$&Case $T=-1$.80\\
\tableline
$\mu_u/\mu_d\equiv T$ & $-$2            & $-$1.80  \\
$\mu_s/\mu_d\equiv U$ & 0.67${}\pm{}$0.03    & 0.70${}\pm{}$0.03    \\
$\mu_{d}^{N}$         & $-$1.25${}\pm{}$0.01 & $-$1.34${}\pm{}$0.01  \\
$\mu_{ d}^{\Lambda}$  & $-$1.10${}\pm{}$0.05 & $-$1.18${}\pm{}$0.05  \\
$\mu_{ d}^{\Sigma}$   & $-$1.13${}\pm{}$0.01 & $-$1.21${}\pm{}$0.01  \\
$\mu_{ d}^{\Xi}$      & $-$1.06${}\pm{}$0.04 & $-$1.13${}\pm{}$0.04  \\
$\mu_d^{\Delta}$      & $-$1.11${}\pm{}$0.05 & $-$1.19${}\pm{}$0.05  \\
$\mu_d^{\Sigma^{*}}$  & $-$1.04${}\pm{}$0.05 & $-$1.11${}\pm{}$0.05  \\
$\mu_d^{\Xi^{*}}$     & $-$0.97${}\pm{}$0.05 & $-$1.04${}\pm{}$0.05  \\
$\mu_d^{\Omega}$      & $-$0.90${}\pm{}$0.05 & $-$0.97${}\pm{}$0.05
\end{tabular}
\end{table}

\mediumtext

\begin{table}
\caption{Our prediction of the decuplet baryon magnetic moments, compared
to the non-relativistic quark model. Most of the errors in the predictions
come from the large error in $\mu(\Delta^{++})$.
In the last column we have used
$\mu(\Delta^{++})=6.0$~$\mu_N$ as an input
corresponding to $\Delta\Sigma(3/2)=0$.
The magnetic moments are given in $\mu_N$.}
\label{tabell2}
\begin{tabular}{ldddd}
&Experiment&NQM&\multicolumn{2}{c}{Our predictions}\\
\tableline

$\mu(\Delta^{++})$ & 4.5${}\pm{}$1.0\protect\cite{bosshand}     & 5.56    &
4.5${}\pm{}$1.0 (input) & 6.0 (input)\\
$\mu(\Delta^{+}) $ &-    & 2.73    & 1.9${}\pm{}$0.6 & 2.8     \\
$\mu(\Delta^{0}) $ &-    & $-$0.09 & $-$0.7${}\pm{}$0.3 & $-$0.3 \\
$\mu(\Delta^{-}) $ &-    & $-$2.92 & $-$3.3${}\pm{}$0.3 & $-$3.5 \\
$\mu(\Sigma^{*+})$ &-    & 3.09    & 2.0${}\pm{}$0.6 & 3.0   \\
$\mu(\Sigma^{*0})$ &-    & 0.27    & $-$0.4${}\pm{}$0.3 & 0.0   \\
$\mu(\Sigma^{*-})$ &-    & $-$2.56 & $-$2.9${}\pm{}$0.2 & $-$3.0  \\
$\mu(\Xi^{*0})$    &-    & 0.63    & $-$0.1${}\pm{}$0.3 & 0.3 \\
$\mu(\Xi^{*-})$    &-    & $-$2.20 & $-$2.4${}\pm{}$0.2 & $-$2.5  \\
$\mu(\Omega^{-})$  & $-$2.024${}\pm{}$0.056\protect\cite{omega} & $-$1.84 &
$-$2.024${}\pm{}$0.056 (input)  &$-$2.024${}\pm{}$0.056 (input)
\end{tabular}
\end{table}

\end{document}